\documentclass[prl,twocolumn,showpacs,superscriptaddress]{revtex4}%
\usepackage{amsfonts,amsmath,amssymb,graphicx,bm,times,color}
\usepackage{amsmath}
\usepackage{amsfonts}
\usepackage{amssymb}
\usepackage{graphicx}%
\providecommand{\U}[1]{\protect\rule{.1in}{.1in}}

\begin{document}
\title{Macroscopic features of quantum fluctuations in large N qubit system}
\author{Andrei~B.~Klimov and Carlos~Mu\~{n}oz }
\affiliation{Departamento de F\'{\i}sica, Universidad de Guadalajara, 44420~Guadalajara,
Jalisco, Mexico}

\begin{abstract}
We introduce a discrete Q-function of N qubit system projected into the space
of symmetric measurements as a tool for analyzing general properties of
quantum systems in the macroscopic limit. For known states the projected
Q-function helps to visualize the results of collective measurements, and for
unknown states it can be approximately reconstructed by measuring lowest
moments of the of collective variables.

\end{abstract}

\date{\today}
\maketitle

\section{Introduction}

It was recently understood that pure states can be used for a statistical
description of quantum system \cite{Popescu}, \cite{Goldstein}. In particular,
the thermal-like pure states \cite{Shimizu} satisfactorily depict equilibrium
values of a certain class of observables \cite{Reimann}. From this point of
view, a quantum state characterization in the limit of macroscopically large
number of particles (macroscopic limit) is tightly connected to the choice of
the observables involved in the measurement process. Since the complexity of
the microscopic description (especially experimentally, through tomographic
reconstruction) of a quantum state grows very rapidly with the dimension of
the system, it is not only very difficult but also unnecessary to determine
individual particle states in order to capture essential properties of large
quantum systems. An adequate choice of measurable coarse-grained \cite{coarse}
collective variables \cite{Hu} assists to reveal several relevant features of
quantum systems in the macroscopic limit. Additionally, in many cases the
assessment of correlation functions of the collective operators are the only
accessible type of practical measurements since it is usually impossible to
distinguish between the particles in a macroscopic volume \cite{colTom}. Thus,
in order to get information about a system as a whole we can restrict
ourselves only to measurement of collective observables symmetric with respect
to particle permutations (symmetric measurements).

In this paper we introduce the concept of the discrete Husimi $Q$-function
projected into the 3 dimensional space of symmetric measurements as a tool for
analysis of the following fundamental questions: How to properly describe a
quantum system state in the macroscopic limit using results of measurements of
appropriate collective variables? How to visualize the result of such
measurements? What higher correlation functions can be described with a
reasonable accuracy by measuring lowest moments of collective observables in
an unknown state?

We will show that the projected $Q$-function is very well suited for
description of $N$-particle systems that behave in many ways as statistical
systems when $N\gg1$, since it naturally emerges from the discrete phase-space
representation and tends to a smooth distribution in the limit of large $N$
preserving essential features of the quantum state. In particular, for known
states it helps to visualize the results of collective measurements, and for
unknown states it is possible to approximately reconstruct the projected
$Q$-function by measuring lowest moments of the collective variables. We
exemplify our approach for the $N$-qubit case; generalization to highest spin
is straightforward.

In Sec.II we briefly recall the concept of the discrete phase-space
distribution functions and propose a discrete $Q$-function in the space of
collective measurements. In Sec.III we analyze the macroscopic limit of this
distribution and introduce the concept of localization in the measurement
space. In Sec. IV we discuss the possibility of reconstructing higher order
moments of collective operators from the asymptotic form of the $Q$%
-distribution. In Sec.V we discuss the results and further topics related to
our approach.

\section{Discrete Q-function in the measurement space}

We first recall that the full non-redundant description of quantum state is
provided by functions in the discrete phase-space. The discrete phase-space
(DPS) for an $N$ qubit system is a two-dimensional $2^{N}\times2^{N}$ grid
where coordinates $(\alpha,\beta)$\ are specified by $N$-dimensional strings
$\alpha=(a_{1},...,a_{N})$, $\beta=(b_{1},...,b_{N})$, $a_{j},b_{j}%
\in\mathbb{Z}_{2}$. The points $(\alpha,\beta)$ label elements of a monomial
operational basis \cite{Schwinger}, \cite{stabilizers} $Z_{\alpha}X_{\beta}$
according to
\[
Z_{\alpha}=\sigma_{z}^{a_{1}}\otimes\ldots\otimes\sigma_{z}^{a_{N}},\quad
X_{\beta}=\sigma_{x}^{b_{1}}\otimes\ldots\otimes\sigma_{x}^{b_{N}},
\]
where $\sigma_{z}=|0\rangle\langle0|-|1\rangle\langle1|$, $\sigma
_{x}=|0\rangle\langle1|+|1\rangle\langle0|$. The operators $Z_{\alpha}$ and
$X_{\beta}$ in the computational basis $\{|\kappa\rangle=|k_{1},...,k_{N}%
\rangle,k_{i}\in\mathbb{Z}_{2}\}$ in the full Hilbert space $\mathcal{H}%
_{2^{N}}$ $=\mathcal{H}_{2}^{\otimes N}$ act as displacements,%
\begin{equation}
Z_{\alpha}|\kappa\rangle=(-1)^{\alpha\kappa}|\kappa\rangle,\quad X_{\beta
}|\kappa\rangle=|\kappa+\beta\rangle, \label{ZX}%
\end{equation}
where the multiplication and sum are \textrm{mod }$2$ operations,
$\alpha\kappa=a_{1}k_{1}+...+a_{N}k_{N}\in\mathbb{Z}_{2}$, $\kappa
+\beta=(b_{1}+k_{1},...,b_{N}+k_{N})$. Defined in this way DPS (isomorphic to
a product of two-dimensional discrete torus $T^{2}\otimes T^{2}\otimes...$) is
endowed with a finite geometry \cite{Wootters87} and admits a set of discrete
symplectic operations required for analysis of quasidistribution functions
\cite{simplect}.

Each point of the DPS also labels an element of a set of the discrete coherent
states (DCS), constructed as \cite{Galetti}
\begin{equation}
|\alpha,\beta\rangle=e^{i\varkappa(\alpha,\beta)}Z_{\alpha}X_{\beta}%
|\xi\rangle\,, \label{discretecs}%
\end{equation}
where $e^{i\varkappa(\alpha,\beta)}$ is an unessential for us phase and
$|\xi\rangle$ is a fiducial state. The set of projectors on (\ref{discretecs})
resolves the identity
\[
\sum_{\alpha,\beta}|\alpha,\beta\rangle\langle\alpha,\beta|=2^{N}%
\mathinner{\hbox{1}\mkern-4mu\hbox{l}}
\]
and constitute a discrete POVM. For instance, the discrete $Q$- and
$P$-symbols of an operator $\hat{f}$ can be formally defined as the following
maps:%
\begin{equation}
Q_{f}{}(\alpha,\beta)=\langle\alpha,\beta|\hat{f}|\alpha,\beta\rangle
,\;\hat{f}=\sum_{\alpha,\beta}P_{f}{}(\alpha,\beta)|\alpha,\beta\rangle
\langle\alpha,\beta|. \label{QP}%
\end{equation}
Nevertheless, not every choice of the fiducial state $|\xi\rangle$ leads to an
informationally complete POVM. It follows from the expression for the mapping
kernel \cite{ruzzi}
\begin{align}
\Delta^{\left(  s\right)  }\left(  \alpha,\beta\right)   &  =\frac{1}{2^{N(s+3)/2}}%
{\displaystyle\sum\limits_{\gamma,\delta}}
\left(  -1\right)  ^{\alpha\delta+\beta\gamma+\gamma\delta(1-s)/2}\notag \\ 
\left[  \left\langle
\xi\right\vert Z_{\gamma}X_{\delta}\left\vert \xi\right\rangle \right]
^{-s} Z_{\gamma}X_{\delta},\label{delta}    \\
Q_{f}\left(  \alpha,\beta\right)   &  =tr\left[  \hat{f}\Delta^{\left(
s=-1\right)  }\left(  \alpha,\beta\right)  \right]  ,\label{dss}\\
P_{f}\left(  \alpha,\beta\right)   &  =tr\left[  \hat{f}\Delta^{\left(
s=1\right)  }\left(  \alpha,\beta\right)  \right]  ,
\end{align}
that the expansion of the density matrix on the DCS projectors is not faithful
(and the $Q$-symbol is not an invertible mapping) for fiducial states where
$\left\langle \xi\right\vert Z_{\gamma}X_{\delta}\left\vert \xi\right\rangle
=0$ for some $\gamma,\delta$.

It was proposed \cite{DCS} to take the fiducial state $|\xi\rangle$ as a
product of identical qubit states:
\[
|\vartheta,\varphi\rangle_{j}=e^{i\varphi/2}\sin\textstyle{\vartheta
/}2|1\rangle_{j}+e^{-i\varphi/2}\cos\textstyle{\vartheta/}2|0\rangle_{j},
\]
with $\vartheta=\arctan\sqrt{2},\varphi=\pi/4$, i.e. $|\xi\rangle$ is the
standard $SU(2)$ coherent state
\begin{equation}
|\vartheta,\varphi\rangle_{1}\otimes\,\ldots\,\otimes|\vartheta,\varphi
\rangle_{N}\propto|\xi=\frac{\sqrt{3}-1}{\sqrt{2}}e^{i\pi/4}\rangle\,,
\label{FS}%
\end{equation}
specified by the unitary vector $\mathbf{n}_{0}=(1,1,1)/\sqrt{3}$. For a
single qubit DCS (\ref{discretecs}) form a symmetric tetrahedron inscribed in
the Bloch sphere, so that $Q$- and $P$-symbols are connected by a simple
relation, $3Q(a,b)=2P(a,b)+1$. In what follows we will refer to DCS with this
choice of the fiducial state as symmetric coherent states \cite{DCS}. Then,
the set of projectors on (\ref{discretecs}) forms a non-orthogonal
informationally complete operational basis. The mappings (\ref{QP}) are
invertible and the average of any operator $\hat{f}$ is computed as a
convolution of the Husimi function $Q_{\rho}(\alpha,\beta)$ with the
$P$-symbol of $\hat{f}$:
\begin{equation}
\langle\hat{f}\rangle=\sum_{\alpha,\beta}P_{f}{}\left(  \alpha,\beta\right)
Q_{\rho}(\alpha,\beta), \label{AV}%
\end{equation}
where $\rho$ is the density matrix.

In the case of systems with continuous symmetries ($H(1)$ for the harmonic
oscillator, $SU(2)$ for a spin-like systems or states from the symmetric
representation of $N$-qubit system, etc.) the quasidistribution functions are
very useful for state identification \cite{WF pic}. Unfortunately, the
representation of states in DPS is not very intuitive \cite{simplect}, mainly
since there is no natural order in the space of two-dimensional $\mathbb{Z}%
_{2}$ strings. The $Q$-function usually has a form of an almost random
distribution of peaks, see Fig.1 where the $Q$-function for the bi-partite
state $\left\vert \psi_{\Gamma}\right\rangle \sim\otimes\Pi_{i=1}^{3}\left[
\left\vert 00\right\rangle _{i}+\left\vert 10\right\rangle _{i}+\left\vert
01\right\rangle _{i}-\left\vert 11\right\rangle _{i}\right]  $ in the 6-qubit
case is plotted.
\begin{figure}
[ptb]
\begin{center}
\includegraphics[
scale=0.25
]%
{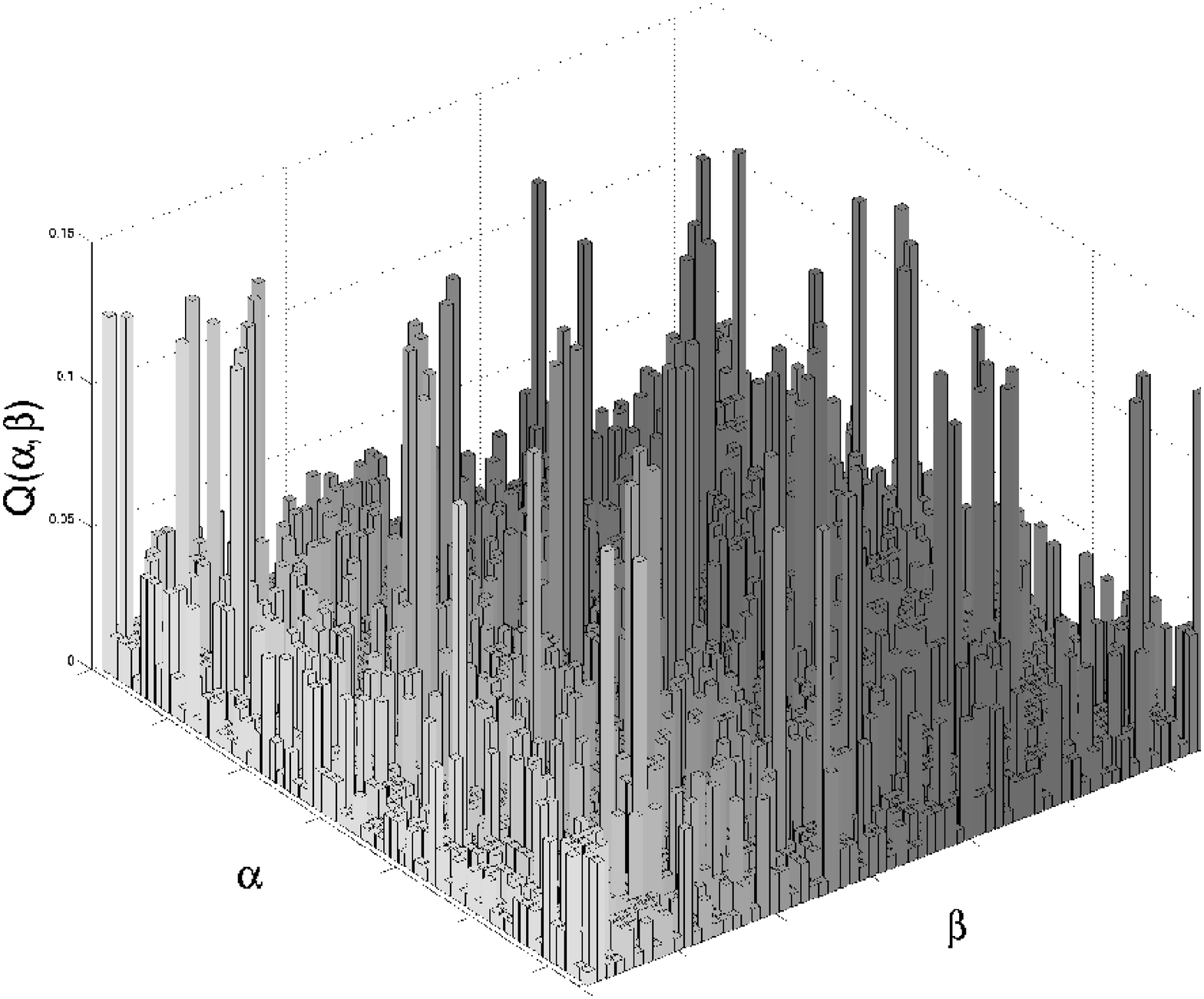}%
\caption{Q-function in the full discrete phase-space, N=6, for the bi-partite
state $\left\vert \psi_{\Gamma}\right\rangle \otimes\Pi_{i=1}^{3}\left[
\left\vert 00\right\rangle _{i}+\left\vert 10\right\rangle _{i}+\left\vert
01\right\rangle _{i}-\left\vert 11\right\rangle _{i}\right]  .$}%
\end{center}
\end{figure}
So, it results very difficult to get a good insight on the quantum state from
the corresponding discrete quasidistribution function especially when the
number of qubits is large, $N>>1$ \cite{DCS}. In addition, the complexity of
the microscopic description of a quantum state grows very rapidly with the
dimension of the system. On the other hand, it is not only very difficult but
also unnecessary to determine individual qubit states in order to describe
essential properties of a quantum system in the macroscopic limit. In order to
describe global properties of $N$-qubit systems we choose the correlation
functions of the collective spin operators
\begin{equation}
S_{x,y,z}=\sum_{i=0}^{N}\,\sigma_{x,y,z}^{(i)} \label{CO}%
\end{equation}
as a set of (symmetric) observables. In practice, average values of powers of
collective operators are non-trivial in any (not only symmetric with respect
to particle permutations) state for except of singlets. Thus, by measuring
symmetric observables (\ref{CO}) we collect information from all $SU(2)$
irreducible subspaces that appear in the decomposition of an arbitrary $N$
qubit state and averaging over hidden degrees of freedom (degeneration of
irreps entering in such decomposition).

The crucial element of our construction is the observation that the $P$ and
$Q$ symbols (\ref{QP}) of symmetric operators do not depend on the phase-space
coordinates but rather only on the lengths of the strings $\alpha,\beta$ and
$\alpha+\beta$ where the length $0\leq h\left(  \kappa\right)  \leq N$ counts
the number of nonzero coefficients $k_{j}\in\mathbb{Z}_{2}$ in the string
$\kappa=(k_{1},k_{2},...,k_{N})$ (in the mathematical literature such
invariant under permutation quantities are usually called weights of the
strings $\alpha,\beta$ and the Humming distance between them), i.e.
\begin{align}
P_{f}\left(  \alpha,\beta\right)   &  =P_{f}\left(  h\left(  \alpha\right)
,h\left(  \beta\right)  ,h(\alpha+\beta)\right)  ,\label{p}\\
Q_{f}\left(  \alpha,\beta\right)   &  =Q_{f}\left(  h\left(  \alpha\right)
,h\left(  \beta\right)  ,h(\alpha+\beta)\right)  . \label{q}%
\end{align}
For instance, one obtains
\begin{align*}
P_{S_{x}}  &  =\frac{\sqrt{3}}{2^{N}}\left[  N-2h\left(  \alpha\right)
\right]  ,\\
P_{S_{y}}  &  =\frac{\sqrt{3}}{2^{N}}\left[  N-2h\left(  \alpha+\beta\right)
\right]  ,\\
P_{S_{z}}  &  =\frac{\sqrt{3}}{2^{N}}\left[  N-2h\left(  \beta\right)
\right]  .
\end{align*}
This feature is specific to the symmetric coherent states (\ref{discretecs}%
)-(\ref{FS}) and follows from the property of (\ref{FS}) that \cite{DCS}
\[
\left\langle \xi\right\vert Z_{\gamma}X_{\delta}\left\vert \xi\right\rangle
=3^{-(h\left(  \gamma\right)  +h\left(  \delta\right)  +h\left(  \gamma
+\delta\right)  )/4}
i^{(h\left(  \gamma\right)  +h\left(  \delta\right)  -h\left(  \gamma
	+\delta\right)  )/2}.
\]
Really, since for any operator $\hat{O}_{c}$ symmetric with respect to
particle permutations it fulfills that $Tr\left[  \hat{O}_{c}Z_{\gamma
}X_{\delta}\right]  =Tr\left[  \hat{O}_{c}Z_{\gamma^{\prime}}X_{\delta
^{\prime}}\right]  $, where $\gamma^{\prime}$ is a permutation of $\gamma$
(i.e. a reordering of elements of the string $\gamma=(\gamma_{1}%
,...,\gamma_{N})$ ) and $\delta^{\prime}$ is \textit{the same} permutation of
$\delta$, then taking into account that $\left(  -1\right)  ^{\alpha
\delta^{\prime}+\beta\gamma\prime}=\left(  -1\right)  ^{\alpha^{\prime}%
\delta+\beta^{\prime}\gamma}$ we obtain that the $P$ and $Q$ symbols
(\ref{dss}), proportional to
\[%
{\displaystyle\sum\limits_{\gamma,\delta}}
\left(  -1\right)  ^{\alpha\delta+\beta\gamma}\left[  \left\langle
\xi\right\vert Z_{\gamma}X_{\delta}\left\vert \xi\right\rangle \right]
^{-s}tr\left[  \hat{O}_{c}Z_{\gamma}X_{\delta}\right]  ,
\]
are invariant under permutations, $Q_{O_{c}}\left(  \alpha,\beta\right)
=Q_{O_{c}}\left(  \alpha^{\prime},\beta^{\prime}\right)  ,P_{O_{c}}\left(
\alpha,\beta\right)  =P_{O_{c}}\left(  \alpha^{\prime},\beta^{\prime}\right)
$ and thus satisfy (\ref{p})-(\ref{q}) (see also Appendix).

The average value (\ref{AV}) of a symmetric observable is then computed as
\begin{align}
\langle\hat{f}\rangle &  =\sum\limits_{m,n=0}^{N}\sum\limits_{k}^{{}}%
P_{f}\left(  m,n,k\right)  \tilde{Q}_{\rho}\left(  m,n,k\right)  ,\\
& \label{f av}\\
k  &  =\left\vert m-n\right\vert ,\left\vert m-n\right\vert +2,...,\min\left(
m+n,N,2N-m-n\right)  ,\nonumber
\end{align}
where $m=h\left(  \alpha\right)  $, $n=h\left(  \beta\right)  $,
$k=h(\alpha+\beta)$ and%

\begin{equation}
\tilde{Q}_{\rho}\left(  m,n,k\right)  =\sum_{\alpha,\beta}Q_{\rho}\left(
\alpha,\beta\right)  \delta_{m,h\left(  \alpha\right)  }\delta_{n,h\left(
\beta\right)  }\delta_{k,h\left(  \alpha+\beta\right)  }, \label{QQ}%
\end{equation}
is the $\tilde{Q}$-function \textit{projected into }$3$\textit{ dimensional
space of symmetric measurements} $\left(  m,n,k\right)  $. The $\tilde
{Q}_{\rho}\left(  m,n,k\right)  $-function contains the entire information
about \textit{all} collective observables. Moreover, the averaging (\ref{QQ})
over strings of given length smooths the original discrete distribution, which
makes it very useful for studying qubit states in the macroscopic limit.

For symmetric states, for which the density matrix is invariant under particle
permutations, the projected $\tilde{Q}$-function according to (\ref{q}) and
(\ref{QQ}) has the form
\begin{equation}
\tilde{Q}_{\rho}\left(  m,n,k\right)  =Q_{\rho}\left(  m,n,k\right)  R_{mnk},
\label{Qsym}%
\end{equation}
where $R_{mnk}$ is the combinatorial factor
\begin{equation}
R_{mnk}=\frac{N!}{\left(  \frac{m+n-k}{2}\right)  !\left(  \frac{2N-m-n-k}%
{2}\right)  !\left(  \frac{n-m+k}{2}\right)  !\left(  \frac{m-n+k}{2}\right)
!}. \label{R}%
\end{equation}
For instance, the $Q$-function of the fiducial state (\ref{FS}) has a
step-like form $Q_{\xi}\left(  \alpha,\beta\right)  =3^{-(h\left(
\alpha\right)  +h\left(  \beta\right)  +h\left(  \alpha+\beta\right)  )/2}$
\cite{DCS}, while the projected $\tilde{Q}$-function (\ref{Qsym})
asymptotically tends to a Gaussian function
\begin{align}
\widetilde{Q}_{\xi}    \sim\exp\left[  -N^{-1}\left(  5/2\left(  m^{2}%
+n^{2}+k^{2}\right)   \right.\right. \notag \\
\left. \left. -km-nk-mn-(m+n+k)N \right)  \right]  ,\label{Q CS}\\
 \nonumber
\end{align}
in the limit $N\gg1$ as a consequence of the smoothing produced by the
combinatorial factor (\ref{R}).

\section{Large N asymptotic and the localization problem}

One can find the asymptotic form of the $\tilde{Q}$-function (\ref{QQ}) in the
macroscopic limit, $N\gg1$, for general states (not necessarily symmetric with
respect to qubit permutations). Making use of an integral representation of
the $\delta$-functions we rewrite (\ref{QQ}) as%
\begin{align}
\tilde{Q}\left(  \mathbf{x}\right)   &  =\frac{1}{i}\int_{|\mathbf{\omega|}%
=1}\frac{f(\mathbf{\omega})}{\omega_{1}^{Nx+1}\omega_{2}^{Ny+1}\omega
_{3}^{Nz+1}}\frac{d^{3}\mathbf{\omega}}{\left(  2\pi\right)  ^{3}}%
\mathbf{,}\label{Qa1}\\
f(\omega)  &  =\sum\limits_{\alpha,\beta}Q\left(  \alpha,\beta\right)
\omega_{1}^{h\left(  \alpha\right)  }\omega_{2}^{h\left(  \alpha+\beta\right)
}\omega_{3}^{h\left(  \beta\right)  },\nonumber
\end{align}
where $\mathbf{x}=(x=m/N,\;y=k/N,\;z=n/N)$ are the scaled coordinates in the
measurement space, and the $P$-symbols of the collective variables
$\mathbf{S\cdot n}$ are
\begin{equation}
P_{\mathbf{S\cdot n}}=\frac{N\sqrt{3}}{2^{N}}\left(  1-2\mathbf{x\cdot
n}\right)  , \label{P1}%
\end{equation}
here $\mathbf{n}$ is a unit vector. If the highest order cumulants $\kappa
_{r}$ of collective variables in any measurement direction do not grow very
rapidly, in the sense that $\kappa_{2}\gg\kappa_{3}^{2/3}$, $\kappa_{2}%
\gg\kappa_{4}^{1/2}$, etc. then applying $\omega_{j}=1+\varepsilon_{j}$,
$|\varepsilon_{j}|\ll1$ we can approximate (\ref{Qa1}) as a single Gaussian
function
\begin{equation}
\tilde{Q}(\mathbf{x)}\approx\frac{2^{N+1}}{(\pi N)^{3/2}\sqrt{\det T}}%
\exp\left(  -N\Delta\mathbf{x}T^{-1}\Delta\mathbf{x}\right)  ,\quad\label{QT}%
\end{equation}
where $\Delta\mathbf{x}=\mathbf{x-\bar{x}}$; here $\mathbf{\bar{x}=}%
(1-\langle\mathbf{S}\rangle/N\sqrt{3})/2$ with%
\begin{equation}
T=\frac{1}{6N}\left(  \Gamma+\Lambda\right)  ,\quad\det T\geq0 \label{T}%
\end{equation}
where
\[
\Gamma_{ij}=\langle S_{i}S_{j}+S_{j}S_{i}\rangle/2-\langle S_{i}\rangle\langle
S_{j}\rangle,\quad i,j=x,y,z
\]
is the correlation matrix and where the symmetric matrix $\Lambda$ is defined
as
\[
\Lambda=\left[
\begin{array}
[c]{ccc}%
2N & \sqrt{3}\left\langle S_{z}\right\rangle  & \sqrt{3}\left\langle
S_{y}\right\rangle \\
\sqrt{3}\left\langle S_{z}\right\rangle  & 2N & \sqrt{3}\left\langle
S_{x}\right\rangle \\
\sqrt{3}\left\langle S_{y}\right\rangle  & \sqrt{3}\left\langle S_{x}%
\right\rangle  & 2N
\end{array}
\right]  ,
\]
if $\det T\neq0$. For instance, for a coherent state $|\mu,\nu\rangle=Z_{\mu
}X_{\nu}|\xi\rangle\,$(not fully symmetric with respect to qubit permutations)
we obtain, considering the covariance of the $Q$-function with respect to
displacements, $Q_{\mu,\nu}\left(  \alpha,\beta\right)  =Q_{\xi}\left(
\alpha+\mu,\beta+\nu\right)  $ \cite{DCS} that $\mathbf{\bar{x}}=(h\left(
\mu\right)  ,h\left(  \mu+\nu\right)  ,h\left(  \nu\right)  )/3N+(1,1,1)/3$
and%
\[
T=\frac{1}{3}\left[
\begin{array}
[c]{ccc}%
4/3 & 1-2\bar{z} & 1-2\bar{y}\\
1-2\bar{z} & 4/3 & 1-2\bar{x}\\
1-2\bar{y} & 1-2\bar{x} & 4/3
\end{array}
\right]  .
\]
It is worth noting that $\tilde{Q}_{\mu,\nu}(\mathbf{x)}$ is totally
isotropic, i.e. is described by a sphere of radius $3N^{-1/2}/2$ centered at
$\mathbf{\bar{x}}=(1,1,1)/2$, $\tilde{Q}_{\mu,\nu}(\mathbf{x)}\sim\exp\left(
-9N\Delta\mathbf{x}^{2}/4\right)  $ only in the case when $h\left(
\mu\right)  =h\left(  \nu\right)  =h\left(  \mu+\nu\right)  =N/2$, which
corresponds to the states with zero average values of the collective spin
operators $S_{x,y,z}$.

Let us point out that the probability ellipsoid $\Delta\mathbf{x}T^{-1}%
\Delta\mathbf{x}$ in (\ref{QT}) depicts how well we can describe the state by
two lowest moments of the collective variables rather than their fluctuations
in some particular directions. For instance, the principal axes of the
ellipsoid (\ref{Q CS}) for the fiducial coherent state (\ref{FS}) are
$\mathbf{n}_{1}=(1,1,1)/\sqrt{3},\mathbf{n}_{2}=(-1,0,1)/\sqrt(2),\mathbf{n}%
_{3}=(-1,1,0)/\sqrt(2)$ with the corresponding lenghts $\lambda_{1}=2/3,\lambda
_{2}=1/3,\lambda_{3}=1/3$. Observe that in the direction $\mathbf{n}_{1}$ the
collective operator $\mathbf{S\cdot n}_{1}$ does not fluctuate at all, since
(\ref{FS}) is its eigenstate. Nevertheless, in this direction the ellipsoid
has maximum length, which means that the accuracy of the state description
with the first and second moments along $\mathbf{n}_{1}$ is less then in the
directions $\mathbf{n}_{2,3}$. In Sec. IV we detail this general assertion
when discuss the higher order moments.

In Fig.2 we plot the exact $\tilde{Q}(\mathbf{x)}$ function for the fiducial
state (\ref{FS}) as a set of spheres in the three dimensional space of
symmetric measurements $\mathbf{\{}m,n,k\}$. The size of the spheres and their
colors indicate the density of the distribution. The largest and darkest
spheres represent points of highest density. The grey envelope corresponds to
the analytical approximation given by (\ref{QT}) centered at $\mathbf{\bar{x}%
}=(1,1,1)/3$.

\begin{figure}[ptb]
\includegraphics[scale=0.35]{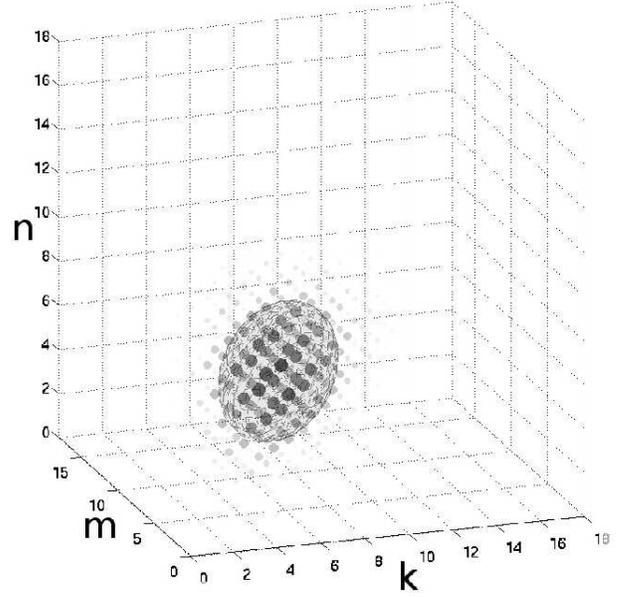}\caption{$\tilde{Q}(\mathbf{x)}$ functions
for an DCS $| 0, 0 \rangle$, with N=18.}%
\end{figure}

In case where one of the eigenvalues of $T$ is zero then along the direction
corresponding to this eigenvalue the $\tilde{Q}$-function is proportional to
the $\delta$-function:
\[
\tilde{Q}(\mathbf{x)}\sim\exp\left(  -N\left(  \mathbf{x}\cdot\mathbf{u}%
_{1}\right)  ^{2}/\lambda_{1}-N\left(  \mathbf{x}\cdot\mathbf{u}_{2}\right)
^{2}/\lambda_{2}\right)  \delta\left(  \mathbf{x}\cdot\mathbf{u}_{3}\right)
,
\]
where $\lambda_{1,2}$ are non zero eigenvalues of $T$, $\lambda_{3}=0$ and
$\mathbf{u}_{j},j=1,2,3$ are the corresponding normalized eigenvectors. An
example of such state is the $SU(2)$ coherent state with $\vartheta
=-\arctan\sqrt{2},\varphi=\pi/4$, for which the $\tilde{Q}(\mathbf{x)}$ is
located in the plane $z=x+y$, so that $\lambda_{1}=\lambda_{2}=2/3,\lambda
_{3}=0$.

The dispersion matrix (\ref{T}) provides important information about state
localization in the space of symmetric measurements. Loosely speaking, the
volume of the ellipsoid $\Delta\mathbf{x}T^{-1}\Delta\mathbf{x}$ is finite for
localized states and becomes unbounded for non-localized states in the limit
$N\rightarrow\infty$. In other words, the state is localized in some direction
if the width of the distribution (\ref{QT}), determined by the eigenvalue of
the matrix $T$ in the corresponding direction, is much less than the extension
of the measurement space. For localized states the outcomes of symmetric
measurements depend only on a small number of parameters, which are
essentially the lowest order moments of the collective variables.

The localization in the measurement space is not the same as localization in
the Hilbert space. For instance, the uniform distribution $\rho=I/2^{N}$ is
represented as a localized sphere\ in the measurement space, $\tilde
{Q}(\mathbf{x)}\sim\exp\left(  -2N\Delta\mathbf{x}^{2}\right)  $ centered at
$\mathbf{\bar{x}}=(1,1,1)/2$.

It follows immediately from (\ref{T}) that the localized states are
characterized by $TrT\ll N$ (so that, $\lim_{N\rightarrow\infty}TrT/N=0$). It
is easy to see that all separable states, $\rho=\otimes\Pi_{i=1}^{N}%
\rho^{\left(  i\right)  }$, are localized since $Tr\Gamma=3N-\sum_{i=1}%
^{N}r_{i}^{2}$, where $r_{i}^{2}$ is the Bloch vector of $i$-th particle, so
that $4/3\leq TrT\leq3/2$. Another examples of localized states are the
bi-separable states of the form
\begin{equation}
\left\vert \psi_{a}\right\rangle \sim\otimes\Pi_{i}\left[  \left\vert
0,1\right\rangle _{i}+a\left\vert 1,0\right\rangle _{i}\right]  , \label{SS}%
\end{equation}
and graph-like states
\begin{equation}
\left\vert \psi_{\Gamma}\right\rangle \sim\otimes\Pi_{i=1}^{N/2}\left[
\left\vert 00\right\rangle _{i}+\left\vert 10\right\rangle _{i}+\left\vert
01\right\rangle _{i}-\left\vert 11\right\rangle _{i}\right]  . \label{GS}%
\end{equation}
For all aforementioned states $\kappa_{r}\sim N$.

One can observe that since for the states (\ref{SS}) the matrix $T$ has the
form
\[
T=\frac{1}{3}diag\left(  \frac{3+3\,{a}^{2}+2\,a}{2\,{a}^{2}+2},\frac
{3+3\,{a}^{2}+2\,a}{2\,{a}^{2}+2},1\right)  ,
\]
the projected $\tilde{Q}$-function for singlet ( $a=-1$) is a sphere,
$\tilde{Q}(\mathbf{x)}\sim\exp\left(  -3N\Delta\mathbf{x}^{2}\right)  $
centered at $\mathbf{\bar{x}}=(1,1,1)/2$. It will be shown below that the
$1/\sqrt{3}$ is the minimum possible radius that a spherically symmetric
distribution can acquire.

For the fully entangled W-state the $T$-matrix in the limit $N\gg1$ has the
form corresponding to the localized states:
\[
T=\frac{1}{6}\left[
\begin{array}
[c]{ccc}%
5 & \sqrt{3} & 0\\
\sqrt{3} & 5 & 0\\
0 & 0 & 2
\end{array}
\right]  .
\]
Nevertheless, taking into account the explicit form of the $Q$-function
\begin{equation}
Q_{W}=\mathcal{N}\left\vert \xi\right\vert ^{2n}\left\vert N\xi+k\left(
\xi^{-1}-\xi\right)  -m\left(  \xi^{-1}+\xi\right)  \right\vert ^{2},
\label{QW}%
\end{equation}
where $\xi=\frac{\sqrt{3}-1}{\sqrt{2}}e^{i\pi/4}$ and $\mathcal{N}%
^{-1}=N(1+\left\vert \xi\right\vert ^{2})^{N}$, it results that the cumulants
in $x$ and $y$ directions behave as $\kappa_{2r}\sim N^{r}$, so that
$\kappa_{2}^{2}\sim\kappa_{4}$ (while along $z$ direction $\kappa_{2r}\sim N$)
and thus do not satisfy the condition required to obtain (\ref{QT}). This is
because the localized $\tilde{Q}_{W}$-function has a fine structure: inside a
single Gaussian envelope centered at $\mathbf{\bar{x}=}1/2(1,1,1-1/\sqrt{3})$
there are two narrow maxima located at
\[
\mathbf{x}_{\pm}\mathbf{\approx}(\frac{1}{2}\pm\frac{\sqrt{3}-1}{2\sqrt{N}%
},\frac{1}{2}\mp\frac{\sqrt{3}-1}{2\sqrt{N}},2-\sqrt{3})
\]
and thus, separated by $\sim N^{-1/2}$ in the $x-y$ plane, as it can be
observed from Fig.3, where the grey envelope corresponds to the Gaussian
approximation (\ref{QT}). The analytical expression for the maxima positions
can be obtained from (\ref{Qsym}), (\ref{R}) and (\ref{QW}) by using Stirling
(but not Gaussian) approximation for $R_{mnk}$.

\begin{figure}[ptb]
\includegraphics[scale=0.35]{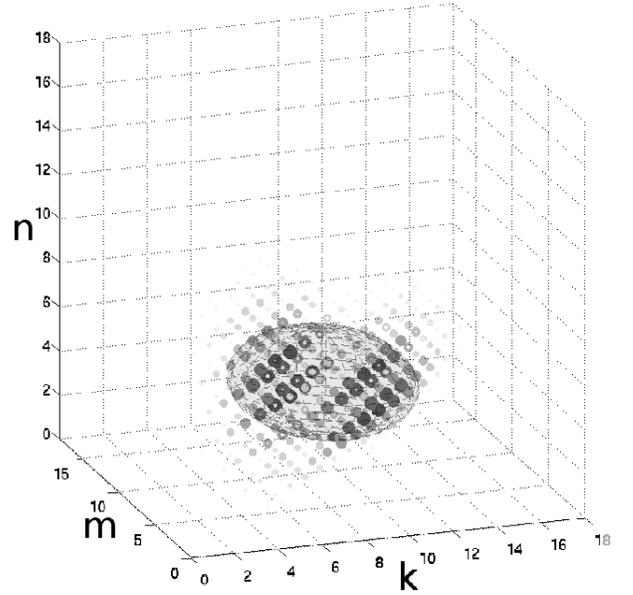}\caption{$\tilde{Q}$-function for the
W-state, N=18.}%
\end{figure}

If at least one of the eigenvalues of the $T$-matrix scales like $N$, the
Gaussian (\ref{QT}) spreads out over the whole measurement space along the
direction corresponding to that eigenvalue. This happens when fluctuations of
the collective variables in some direction behave like $\sim N^{2}$. Such
non-localized states can not be described in the measurement space by a single
Gaussian. Consider as an example the state $\left\vert GHZ\right\rangle
=\left[  \left\vert 0...0\right\rangle +\left\vert 1...1\right\rangle \right]
/\sqrt{2}$. The correlation matrix has the form $\Gamma=diag(N,N,N^{2})$, so
taking into account that $\left\langle S_{x,y,z}\right\rangle =0$ we obtain
from (\ref{QT}) the following Gaussian approximation for the $\tilde{Q}$-function%

\begin{equation}
\tilde{Q}(\mathbf{x)}\sim\exp\left[  -2N\left(  \Delta x^{2}+\Delta
y^{2}+3\Delta z^{2}/N\right)  \right]  , \label{Q GHZ G}%
\end{equation}
where $\mathbf{\bar{x}}=(1,1,1)/2$: the $\tilde{Q}$-function is delocalized
along the axis $z$ in the measurement space, which also follows from a rapidly
growing behavior of cumulants, $\kappa_{2r}\sim N^{2r}$ along the $z $-direction.

In Fig.4 we plot the exact form (\ref{Qsym}) of the $\tilde{Q}$-function for
the GHZ state projected into the space of symmetric measurements. Observe that
along the "localized directions" $x$ and $y$ the $\tilde{Q}$-function is well
described the Gaussian (\ref{Q GHZ G}). The grey envelope corresponds to the
approximation (\ref{QT}) and although does not describe well the general
behavior, still provides non-trivial bounds for the distribution. Using the
explicit form of the discrete $Q$-function for the GHZ state
\begin{equation}
Q_{GHZ}=\mathcal{N}N\left\vert \xi^{n}+\left(  -1\right)  ^{m}\xi
^{N-n}\right\vert ^{2}/2, \label{Q GHZ}%
\end{equation}
one can find from (\ref{Qsym}) and (\ref{Q GHZ}) that the projected to the
measurement space $\tilde{Q}$-function is described by a sum of two equally
weighted localized Gaussians centered at $(1,1,1\pm1/\sqrt{3})/2$ with
corresponding dispersion matrices:
\begin{equation}
T_{\pm}=\left[
\begin{array}
[c]{ccc}%
3 & \pm\sqrt{3} & 0\\
\pm\sqrt{3} & 3 & 0\\
0 & 0 & 3
\end{array}
\right]  . \label{Tpm}%
\end{equation}

\begin{figure}[ptb]
\includegraphics[scale=0.35]{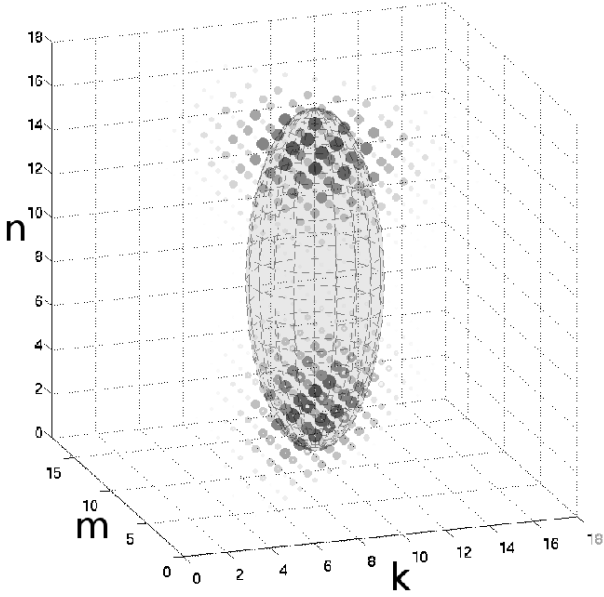}\caption{$\tilde{Q}$-function for the GHZ
state, N=18.}%
\end{figure}Another example of non-localized states is a superposition of
elements of the computational basis $|\kappa\rangle$: $|\psi\rangle\sim
|\kappa_{1}\rangle+|\kappa_{2}\rangle$ with $h(\kappa_{1}+\kappa_{2}%
)\gtrsim\sqrt{N}$; this should be contrasted with the case of the single
$|\kappa\rangle$ for which the $\tilde{Q}$-function is a localized ellipsoid
centered at $(1/2,1/2,1/2+(h(\kappa)/N-1/2)3^{-1/2})$ and characterized by
\begin{equation}
T=\frac{1}{2}\left[
\begin{array}
[c]{ccc}%
1 & (1-2h(\kappa)/N)/\sqrt{3} & 0\\
(1-2h(\kappa)/N)/\sqrt{3} & 1 & 0\\
0 & 0 & 2/3
\end{array}
\right]  . \label{Tk}%
\end{equation}

Localization and delocalization properties are in general not invariant under
local transformations generated by (\ref{ZX}). For instance, the correlation
matrix for the shifted GHZ state,
\[
X_{\nu}\left\vert GHZ\right\rangle \sim\left\vert \nu_{1},\nu_{2}%
,...\right\rangle +\left\vert 1+\nu_{1},1+\nu_{2},...\right\rangle ,
\]
has the form $\Gamma=diag(N,N,\left(  N-2h\left(  \nu\right)  \right)  ^{2})$,
$\nu=(\nu_{1},...,\nu_{N})$, $\nu_{j}=0,1$. Thus, for translations
characterized by lengths $h\left(  \nu\right)  \sim N/2\pm\Delta h$, with
$\Delta h\sim\sqrt{N}$ the states $X_{\nu}\left\vert GHZ\right\rangle $ become
localized and can be sufficiently well described by a single Gaussian envelope
(\ref{QT}). In Fig.5 we plot the $\tilde{Q}$-function for the shifted GHZ
state with $\nu=(1,1,1,1,0,0,0,0)$ in 8 qubit case. The analytical
approximation corresponds to an ellipsoid described by $T=diag(1/2,1/2,1/3)$.

\begin{figure}[ptb]
\includegraphics[scale=0.35]{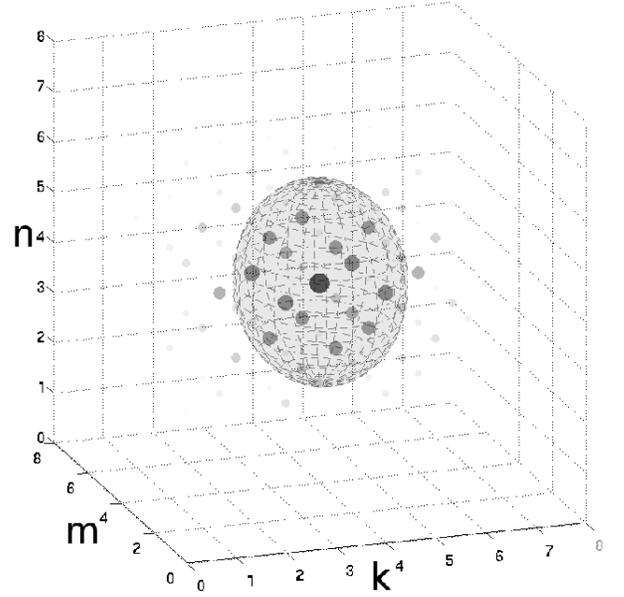}\caption{$\tilde{Q}$-function for the
state $\left\vert 11110000\right\rangle +\left\vert 00001111\right\rangle $,
N=8.}%
\end{figure}On the other hand the states (\ref{SS}) and (\ref{GS}) remain
localized under transformations (\ref{ZX}).

Thus, analyzing the dispersion matrix (\ref{T}) obtained from measured data we
can assess the localization properties of unknown states: unbounded for
$N\rightarrow\infty$ scaled eigenvalues $\lambda_{j}$ of $T$, such that
$\lim_{N\rightarrow\infty}(\lambda_{j}/N)=\infty$, determine unlocalized
directions in the measurement space, while in the direction of finite
eigenvalues $\lambda_{j}$ the distribution function of measured collective
observables is delimited by a narrow Gaussian (\ref{QT}).

\section{Average value evaluation}

The Gaussian approximation (\ref{QT}) is not only helpful to describe the
shape of the $\tilde{Q}$-function in the large $N$ limit, but can be also used
to estimate the highest moments of the collective observables by integrating
(\ref{QT}) with $P(\mathbf{x)}$-symbols of the corresponding variables\ %

\begin{equation}
\left\langle \hat{O}\right\rangle =\frac{N^{3}}{2}%
{\displaystyle\iiint}
P_{O}(\mathbf{x)}\tilde{Q}\left(  \mathbf{x}\right)  d^{3}x, \label{avv}%
\end{equation}
where $\hat{O}$ is an arbitrary symmetric operator and $1/2$ factor appears
because the summation index $k$ in (\ref{f av}) runs in steps of two. The
approximate $\tilde{Q}$-function (\ref{QT}) leads to exact expressions for the
first and second moments of collective observables.

In the case of localized states this approximation also satisfactorily
describes the leading-term asymptotic of the higher order correlations. For
instance, for the shifted coherent states $Z_{\mu}X_{\nu}|\xi\rangle$ the
deviation of the approximate expressions (\ref{avv}) from the exact ones for
the averages $\langle S_{zc}{}^{3}\rangle$, $\langle S_{zc}{}^{4}\rangle$ of
the central moments $S_{jc}=S_{j}-\langle S_{j}{}\rangle$ appears in the
orders $N$ and $N^{2}$ correspondingly:%
\begin{align*}
\langle S_{zc}{}^{3}\rangle_{ex}  &  =-3^{-3/2}Nc_{\nu}\left[  N^{2}c_{\nu
}+6N-4\right]  ,\\
\langle S_{zc}{}^{3}\rangle_{app}  &  \approx-3^{-3/2}Nc_{\nu}\left[
N^{2}c_{\nu}+6N+12\right]  ,
\end{align*}

\begin{align*}
\langle S_{zc}{}^{4}\rangle_{ex}  &  =\frac{1}{9}N^{2}\left(  N^{2}c_{\nu}%
^{4}+\left[  12N-16\right]  c_{\nu}{}^{2}+12\right)  ,\\
\langle S_{zc}{}^{4}\rangle_{app}  &  \approx\frac{1}{9}N^{2}\left(
N^{2}c_{\nu}{}^{4}+\left[  12N+48\right]  c_{\nu}{}^{2}+12+\frac{96}%
{N}\right)  ,
\end{align*}
where $c_{\nu}=2h\left(  \nu\right)  /N-1$. Similar expressions are obtained
for $S_{xc}$ and $S_{yc}$ (just changing the index $\nu$ to $\mu$ and $\nu
+\mu$ correspondingly). Thus, the relative deviation is $\sim N^{-2}$ for
non-zero values of the shifting coefficients $c_{j},j=\nu$, $\mu$, $\nu+\mu$.
It is interesting to note that even for the deeply non-symmetric case, when
all $c_{j}=0$ and the projected $\tilde{Q}$-function takes a form of a ball in
the measurement space, the Gaussian approximation (\ref{QT}) provides good
results for the higher-order moments. In the displaced GHZ state $X_{\nu
}\left\vert GHZ\right\rangle $, with $h\left(  \nu\right)  =N/2$, described by
a localized distribution, such relative deviation is $\sim N^{-1}$.

For non-localized states the approximation (\ref{QT}) does not lead to an
accurate estimate of highest moments along the directions where the
distribution is spread in the measurement space. For instance, the relative
deviation of the forth order moments in the non-localized $x$-direction is
$\sim1$ for the $W$-state ,%

\begin{align*}
\left\langle S_{xc}^{4}\right\rangle _{app}  &  \approx27N^{2}+12N-20,\\
\;\left\langle S_{xc}^{4}\right\rangle _{ex}  &  =15N^{2}-30N+16,
\end{align*}
while in the localized $z$-direction it is still $\sim N^{-2}$:%

\begin{align*}
\langle S_{zc}^{4}{}\rangle_{ex}  &  =\left[  N-2\right]  ^{4},\\
\langle S_{zc}^{4}{}\rangle_{app}  &  \approx\left[  N-2\right]
^{4}+16\left[  N-2\right]  ^{2}.
\end{align*}
A similar situation occurs in the GHZ state if the approximation
(\ref{Q GHZ G}) is used:%
\[
\left\langle S_{xc}^{4}\right\rangle _{app}\approx3N^{2}+16N,\;\left\langle
S_{xc}^{4}\right\rangle _{ex}=3N^{2}-2N,
\]

\[
\left\langle S_{zc}^{4}{}\right\rangle _{app}\approx3N^{4}+16N^{2}%
,\;\left\langle S_{zc}^{4}{}\right\rangle _{ex}=N^{4}.
\]
The two-Gaussian approximation (\ref{Tpm}) provides a correct estimate for the
leading terms of highest moments even in the non-localized direction with a
relative deviation $\sim N^{-2}$: $\left\langle S_{zc}^{4}{}\right\rangle
\approx N^{4}+16N^{2}$.

As was mentioned in Sec.II, the accuracy of estimation of higher order moments
from the two lowest ones for localized states can be related to the lengths of
principal axes. In general, the larger is the length of the axis the less is
the accuracy of estimation along this direction (for appropriately normalized
operators). For instance, for the fiducial state (\ref{FS}) the average value
$\langle(\mathbf{S\cdot n_{1})}^{k}\rangle=N^{k}$, $\mathbf{n}_{1}%
=(1,1,1)/\sqrt{3}$ can not be obtained as a result of integration of the
Gaussian (\ref{Q CS}) with the corresponding $P$-function for $k>4$. As a
representative additional example, let us consider an element of computational
basis $|\kappa\rangle$, for which the T-matrix has the form (\ref{Tk}) and
thus the principal axes and the corresponding eigenvalues are%

\begin{align*}
\mathbf{n}_{1}  &  =(1,0,1)/\sqrt{2},\quad\mathbf{n}_{2}=(-1,0,1)/\sqrt
{2},\quad\mathbf{n}_{3}=(0,0,1),\\
\lambda_{1}  &  =\frac{1}{2}+\frac{\sqrt{3}}{6}\left[  1-2\gamma\right]
,\;\lambda_{2}=\frac{1}{2}-\frac{\sqrt{3}}{6}\left[  1-2\gamma\right]
,\;\lambda_{3}=\frac{1}{3},
\end{align*}
where $\gamma=h\left(  \kappa\right)  /N.$ The exact and approximate values of
the forth order moments along $\mathbf{n}_{1}$ and $\mathbf{n}_{3}$ are then
\begin{align*}
\langle S_{zc}^{4}{}\rangle_{ex}  &  =\left[  1-2\gamma\right]  ^{4}N^{4},\\
\langle S_{zc}^{4}{}\rangle_{app}  &  \approx\left[  1-2\gamma\right]
^{4}N^{4}+16\left[  1-2\gamma\right]  ^{2}N^{2},\\
\langle\left(  S_{xc}+S_{yc}\right)  ^{4}{}\rangle_{ex}  &  =12N^{2}-8N,\\
\langle\left(  S_{xc}+S_{yc}\right)  ^{4}{}\rangle_{app}  &  \approx
12N^{2}+\left(  208+48\sqrt{3}-96\sqrt{3}\gamma\right)  N.
\end{align*}
In order to analyze these results we first observe that the average value
$\langle S_{zc}{}\rangle_{ex}=\left[  1-2\gamma\right]  N$, while $\langle
S_{xc}+S_{yc}\rangle_{ex}=0$, thus appropriately normalized deviations of
approximate moments from the exact ones are
\begin{align}
N^{2}\frac{\langle S_{zc}^{4}{}\rangle_{app}-\langle S_{zc}^{4}{}\rangle_{ex}%
}{\langle S_{zc}^{4}{}\rangle_{ex}}  &  =\frac{16}{\left[  1-2\gamma\right]
^{2}},\label{z dev}\\
N\frac{\langle\left(  S_{xc}+S_{yc}\right)  ^{4}{}\rangle_{app}-\langle\left(
S_{xc}+S_{yc}\right)  ^{4}{}\rangle_{ex}}{\langle\left(  S_{xc}+S_{yc}\right)
^{4}{}\rangle_{ex}}  &  =18+4\sqrt{3}(1-2\gamma). \label{x dev}%
\end{align}
It follows form this that for the state $|0\rangle$ (all qubits are
non-excited, $\gamma=0$) $\lambda_{1}>\lambda_{3\text{ }}$and the deviation
(\ref{x dev}) is larger than (\ref{z dev}), while for $|1\rangle$ (all qubits
non-excited, $\gamma=1$) $\lambda_{3}>\lambda_{1\text{ }}$and the situation is
inverse. The case $\gamma=1/2$ corresponding to a half excited qubits is
special, since in such state $\langle S_{zc}^{m}{}\rangle=0$ for all $m$.

\textit{Spherically symmetric localized distributions}. It is clear that the
collective variables $\left(  \mathbf{S\cdot n}_{j}\right)  ,j=1,2,3$ aligned
along the principal axes of the probability ellipsoid (\ref{QT})
$\Delta\mathbf{x}T^{-1}\Delta\mathbf{x}$ have zero average values. Thus, it
follows from the form of the P-function for the first order moments (\ref{P1})
that a spherically symmetric distribution can be centered only at
$(1/2,1/2,1/2)$, which implies that all the average values $\left\langle
S_{x,y,z}\right\rangle $ are automatically zero. On the other hand, for a
spherical distribution $\tilde{Q}(\mathbf{x)}\sim\exp\left(  -Nr\Delta
\mathbf{x}^{2}\right)  $, characterized by a single parameter $r$, the average
value
\[
\left\langle S_{j}^{2}\right\rangle =2N\frac{3-r}{r},
\]
which impose a restriction $r\leq3$, where the equality is reached at singlet
states. It is worth recalling here that the parameter $r=2$ corresponds to the
uniform state $2^{-N}I$.

\section{Conclusions}

In summary, discrete quasidistribution functions projected into the space of
symmetric measurements provide useful insight into the macroscopic behavior of
large particle systems. They can be used both for visualization purposes and
for the analysis of general properties\ of quantum states from measured data
in the asymptotic limit $N\rightarrow\infty$. In particular, it is suitable
for testing the localization of $N$-qubit states and for estimation of the
higher order correlations of collective observables. The shape of the
projected $\tilde{Q}$-function characterize the precision of the state
description with the lowest moments. In particular, the first two moments of
the collective variables in the localized directions contain important
information about general properties of the state (e.g. allow to estimate the
higher order correlations).

We have numerically tested several types of localized states and for all of
them the Gaussian approximation (\ref{QT}) gives exact coefficient of the
leading term for \textit{all} higher moments. In other words, the higher
moments of normalized collective operators $\mathbf{S\cdot n/}N$ are
completely determined by the first two moments. In this sense some of the localized
states in the measurement space are similar of the 3 dimensional harmonic
oscillator coherent states.

Especially interesting in this respect are spherically symmetric localized
states. Due to the statistical independence of all directions the average
value computed according to (\ref{avv}) of the commutator of any operator of
the form $(\mathbf{S\cdot n)}^{k}$ (and thus an arbitrary element from the
enveloping algebra of collective operators) with $\mathbf{S\cdot n}%
\acute{}%
$ over spherically symmetric states is zero to the leading order for all
$\mathbf{n,n%
\acute{}%
}$. Thus, the operators $\mathbf{S\cdot n}$ can be effectively considered as
classical macroscopic observables on the symmetric states.

Some of non-localized states (GHZ states, etc) can be represented in
the measurement space as a superposition of Gaussians (\ref{QT}), which
provides a transparent physical picture about sets of measurements that allow
to detect such states. For instance, the GHZ state is well described by two
Gaussians (\ref{Tpm}), the assessment of which requires measurements of the
fluctuations of central moments $S_{jc}=S_{j}-\langle S_{j}{}\rangle$ at
$\langle\vec{S}_{\pm}\rangle=(0,0,\pm N)$ along directions of determined by
the principal axes: $\mathbf{n}_{1}=(1,-1,0)/\sqrt{2},\mathbf{n}%
_{2}=(1,1,0)/\sqrt{2},\mathbf{n}_{3}=(0,0,1)$.

Interestingly, small subspaces (of the whole $2^{N}$ dim Hilbert space) with
a fixed energy do not necessarily correspond to the localized states. For
instance, the uniform state in a subspace with a fixed eigenvalue $N(N+2)$ of
the Hamiltonian $H=S_{x}^{2}+S_{y}^{2}+S_{z}^{2}$, described by
\[
\rho=N^{-1}\sum_{k=-N/2}^{N/2}|k,N/2\rangle\langle k,N/2|,
\]
where $|k,N/2\rangle$ is a completely symmetric under permutation $N$ particle
state with $S_{z}|k,N/2\rangle=(2k-N)|k,N/2\rangle$ (the Dicke state), is not
localized in the measurement space. The relation of our description of large
qubit systems to the thermodynamical approach through a Hamiltonian evolution
is an intriguing problem and will be considered elsewhere \cite{Hmes}.

The same approach can be extended to systems with higher symmetries (qudits),
in order to analyze the difference between large numbers of two and many-level
systems in the macroscopic limit.

Similarly to the projected $\tilde{Q}$-function it could be expected that a
projected into the space of symmetric measurements discrete Wigner function
$W_{\rho}(\alpha,\beta)$ would be an interesting tool for observation of
quantum interference in the macroscopic limit. Nevertheless, there is an
important difference between the Wigner and $Q$-functions on the geometrical
level: it is usually expected that summing the Wigner functions along
appropriate directions in the discrete phase-space (rays, for the standard
construction) one obtains the probability to detect the systems in the states
associated to such lines \cite{Wootters87}. In order to satisfy such property
one should insert a phase $\phi(\gamma,\delta)$ into the kernel (\ref{delta})
and set $s=0$. Such a phase satisfies the equation $\phi^{2}(\gamma
,\delta)=(-1)^{\gamma\delta}$ (for the standard construction \cite{Wootters87}%
) and thus, is not uniquely defined \cite{simplect}. On the other hand, since
the Wigner mapping is self-dual, i.e.
\[
\langle\hat{f}\rangle=\sum_{\alpha,\beta}W_{f}{}\left(  \alpha,\beta\right)
W_{\rho}(\alpha,\beta),
\]
a meaningful projected Wigner function can be defined only if the symbols of
collective operators $W_{f}{}\left(  \alpha,\beta\right)  $ are functions of
corresponding lengths, as it happens for $Q$ and $P$-functions (\ref{q}%
)-(\ref{p}). This requirement establishes an additional non-trivial condition
of the phase $\phi(\gamma,\delta)$. It would be interesting to find a phase
that satisfies such a condition and simultaneously guarantee the self-duality
and the geometrical properties of the Wigner mapping.

We would like to thank Prof. H. de Guise for stimulating discussions and
useful remarks.

This work is supported by the Grant 106525 CONACyT, Mexico.

\bigskip

\section{Appendix}

In this Appendix we present some explicit expressions for computation of
symbols of collective operators.

The generating function for the $Q$-function of the moments of the collective
operator in an arbitary direction $\mathbf{S\cdot n}$ has the form%
\begin{gather*}
Q(\alpha,\beta)=\left\langle \alpha,\beta\right\vert \exp\left(
\lambda\mathbf{S\cdot n}\right)  |\alpha,\beta\rangle\\
=\left(  -1\right)  ^{h\left(  \alpha\right)  }\left[  \frac{\sinh\left(
\lambda\right)  }{\sqrt{3}}\right]  ^{N}\\
\left[  n_{x}+n_{y}+n_{z}+\sqrt{3}\coth\left(  \lambda\right)  \right]
^{N-\frac{h\left(  \alpha\right)  +h\left(  \beta\right)  +h\left(
\alpha+\beta\right)  }{2}}\\
\left[  n_{x}+n_{y}-n_{z}-\sqrt{3}\coth\left(  \lambda\right)  \right]
^{\frac{h\left(  \alpha\right)  -h\left(  \beta\right)  +h\left(  \alpha
+\beta\right)  }{2}}\\
\left[  n_{x}-n_{y}-n_{z}+\sqrt{3}\coth\left(  \lambda\right)  \right]
^{\frac{-h\left(  \alpha\right)  +h\left(  \beta\right)  +h\left(
\alpha+\beta\right)  }{2}}\\
\left[  n_{x}-n_{y}+n_{z}-\sqrt{3}\coth\left(  \lambda\right)  \right]
^{\frac{h\left(  \alpha\right)  +h\left(  \beta\right)  -h\left(  \alpha
+\beta\right)  }{2}}.
\end{gather*}
The $P$-function has the same symmetry properties, but its computation is more
involved and requires evaluation of the following sums:%
\begin{align*}
P\left(  \alpha,\beta\right)    & =\frac{1}{2^{3N}}%
{\displaystyle\sum\limits_{\gamma,\delta}}
(-1)^{\alpha\delta+\beta\gamma}3^{h\left(  \gamma\right)  +h\left(
\delta\right)  +h\left(  \gamma+\delta\right)  }\\
&
{\displaystyle\sum\limits_{\mu,\nu}}
(-1)^{\mu\delta+\nu\gamma}Q\left(  \mu,\nu\right),
\end{align*}
For instance, for the powers of $S_{x}$ one obtaines
\[
P_{S_{x}^{2}}=3\,{\frac{\left(  2\,x-1\right)  ^{2}{N}^{2}}{{2}^{N}}}%
-\,{\frac{N}{{2}^{N-1}}},%
\]

\begin{align*}
P_{S_{x}^{3}}  &  =-\,{\frac{3^{3/2}\left(  2\,x-1\right)  ^{3}{N}^{3}}%
{{2}^{N}}}+{\frac{3^{3/2}\left(  2\,x-1\right)  {N}^{2}}{{2}^{N-1}}}\\
&  -{\frac{\sqrt{3}N\left(  2\,x-1\right)  }{{2}^{N-2}}},%
\end{align*}

\begin{align*}
P_{S_{x}^{4}}  &  =9\,{\frac{\left(  2\,x-1\right)  ^{4}{N}^{4}}{{2}^{N}}%
}-36\,{\frac{\left(  2\,x-1\right)  ^{2}{N}^{3}}{{2}^{N}}}\\
&  +12\,{\frac{\left(  16\,{x}^{2}-16\,x+5\right)  {N}^{2}}{{2}^{N}}%
}-32\,{\frac{N}{{2}^{N}},}%
\end{align*}
where $x=m/N$.


\begin{thebibliography}{99}                                                                                               %


\bibitem {Popescu}S. Popescu, A.J. Short, and A. Winter, Nature Phys.
\textbf{2}, 754 (2006).

\bibitem {Goldstein}S. Goldstein , J.L. Lebowitz, R. Tumulka, N. Zanghi, Phys.
Rev. Lett. \textbf{96}, 050403 (2006).

\bibitem {Shimizu}S. Sugiura and A. Shimizu, Phys. Rev. Lett. \textbf{108}, 240401(2012).

\bibitem {Reimann}P. Reimann, Phys. Rev. Lett. \textbf{99}, 160404 (2007).

\bibitem {coarse}J. von Neumann, Z. Phys. \textbf{57}, 30 (1929); S.
Goldstein, J.L. Lebowitz, C. Mastrodonato, R. Tumulka, and N. Zanghi, Phys.
Rev.E \textbf{81}, 011109 (2010).

\bibitem {Hu}C.H. Chou, B. Hu and T. Yu, Physica A \textbf{387} 432 (2008);
B.L. Hu and Y. Subasi, arXiv:1304.7839v1 (2013).

\bibitem {colTom}G. T\'{o}th, W. Wieczorek, D. Gross, R. Krischek, C.
Schwemmer, and H. Weinfurter, Phys. Rev. Lett. \textbf{105}, 250403 (2010); T.
Moroder, P. Hyllus, G. T\'{o}th, C. Schwemmer, A. Niggebaum, S. Gaile, O.
G\"{u}hne, and H. Weinfurter, New J. Phys. \textbf{14} 105001 (2012).

\bibitem {Schwinger}J. Schwinger , Proc. Natl. Acad. Sci. USA \textbf{46}, 570
(1960); ibid. \textbf{46}, 883 (1960); ibid. \textbf{46}, 1401 (1960).

\bibitem {stabilizers}D. Gottesman, Phys. Rev. A \textbf{54}, 1862 (1996); E.
Hostens, J. Dehaene, and B. De Moor, Phys. Rev. A \textbf{71}, 042315 (2005).

\bibitem {Wootters87}W.K. Wootters, Ann. Phys. \textbf{176}, 1 (1987); K.S.
Gibbons, M.J. Hoffman, and W.K. Wootters, Phys. Rev. A \textbf{70}, 062101 (2004)

\bibitem {simplect}A. Vourdas, Rep. Prog. Phys. \textbf{67} 267 (2004); J.P.
Paz, A. J. Roncaglia, and M. Saraceno, Phys. Rev. A \textbf{72}, 012309
(2005); C. Cormick, E.F. Galvao, D. Gottesman, J.P. Paz, and A.O. Pittenger
Phys. Rev. A, 73, 012301 (2006); A.~B. Klimov, C. Mu{\~{n}}oz and J.~L.
Romero, J. Phys. A \textbf{39}, 14471 (2006); G. Bjork, A.B. Klimov and L.L.
Sanchez-Soto Prog. Opt. \textbf{51} 470 (2008).

\bibitem {ruzzi}M. Ruzzi, M. A. Marchiolli, and D. Galetti, J. Phys. A
\textbf{38}, 6239 (2005); A.B. Klimov, C. Munoz, and L.L. Sanchez-Soto, Phys.
Rev. A \textbf{80}, 043836 (2009 ).

\bibitem {Galetti}D. Galetti, and M.A. Marchiolli, Ann. Phys. \textbf{249},
454 (1996).

\bibitem {DCS}C. Mu{\~{n}}oz, A.B. Klimov, and L.L. Sanchez-Soto, J. Phys. A.
\textbf{45}, 244014 (2012); A.B. Klimov and C. Mu{\~{n}}oz, Phys. Scr.
\textbf{87} 038110 (2013).

\bibitem {WF pic}J.P. Dowling, G.S. Agarwal, and W.P. Schleich, Phys. Rev. A
\textbf{49}, 4101 (1994).

\bibitem {Hmes}C. Mu{\~{n}}oz, M. Gaeta, and A.B. Klimov, in preparation.
\end{thebibliography}
\end{document}